\documentclass[aps,prx,showpacs,twocolumn,amsmath,amssymb,superscriptaddress,letterpaper]{revtex4}
\usepackage{times}
\usepackage{amsfonts}
\usepackage{mathrsfs}
\usepackage{graphicx}
\usepackage{dcolumn}
\usepackage{bm}
\usepackage{color}

\usepackage[colorlinks,bookmarks=false,citecolor=blue,linkcolor=red,urlcolor=blue]{hyperref}
\def\be{\begin{equation}}       \def\ee{\end{equation}}
\def\bea{\begin{eqnarray}}      \def\eea{\end{eqnarray}}

\begin{document}

\title{Topological Vortex Phase Transitions  in Iron-Based Superconductors }
\author{Shengshan Qin}
\affiliation{Kavli Institute of Theoretical Sciences, University of Chinese Academy of Sciences, Beijing, 100049, China}
\affiliation{Beijing National Laboratory for Condensed Matter Physics, Institute of Physics, Chinese Academy of Sciences, Beijing 100190, China}

\author{Lunhui Hu}
\affiliation{Department of Physics, University of California, San Diego, California 92093, USA}
\affiliation{Kavli Institute of Theoretical Sciences, University of Chinese Academy of Sciences, Beijing, 100049, China}

\author{Xianxin Wu}
\affiliation{Institute for Theoretical Physics and Astrophysics, Julius-Maximilians University of W\"urzburg, Am Hubland, D-97074 W\"urzburg, Germany}

\author{Xia Dai}
\affiliation{Key Laboratory of Aperture Array and Space Application, Hefei, 230088, China}


\author{Chen Fang}
\affiliation{Beijing National Laboratory for Condensed Matter Physics, Institute of Physics, Chinese Academy of Sciences, Beijing 100190, China}
\affiliation{CAS Center for Excellence in Topological Quantum Computation, University of Chinese Academy of Sciences, Beijing, 100049, China}

\author{Fu-chun Zhang}\email{fuchun@ucas.ac.cn}
\affiliation{Kavli Institute of Theoretical Sciences, University of Chinese Academy of Sciences, Beijing, 100049, China}
\affiliation{CAS Center for Excellence in Topological Quantum Computation, University of Chinese Academy of Sciences, Beijing, 100049, China}

\author{Jiangping Hu}\email{jphu@iphy.ac.cn}
\affiliation{Beijing National Laboratory for Condensed Matter Physics, Institute of Physics, Chinese Academy of Sciences, Beijing 100190, China}
\affiliation{Kavli Institute of Theoretical Sciences, University of Chinese Academy of Sciences, Beijing, 100049, China}
\affiliation{CAS Center for Excellence in Topological Quantum Computation, University of Chinese Academy of Sciences, Beijing, 100049, China}
\affiliation{Collaborative Innovation Center of Quantum Matter, Beijing, China}

\date{\today}

\begin{abstract}
We study topological vortex phases in iron-based superconductors. Besides the previously known vortex end Majorana zero modes (MZMs) phase stemming from the existence of a three dimensional (3D) strong topological insulator state, we show that there is another topologically nontrivial phase as iron-based superconductors can be doped superconducting 3D weak topological insulators (WTIs).  The vortex bound states in a superconducting 3D WTI exhibit two different types of quantum states, a robust nodal superconducting phase with pairs of bulk MZMs and a full-gap topologically nontrivial superconducting phase which has single vortex end MZM in a certain range of doping level. Moreover, we predict and summarize various topological phases in iron-based superconductors, and find that carrier doping and interlayer coupling  can drive systems to have phase transitions between these different topological phases.

\end{abstract}


\maketitle

\section {\bf introduction}

\emph{Introduction}.---For the possible application in quantum computation, the search for Majorana zero modes (MZMs) has been one of the most intriguing topics in condensed matter physics\cite{quantum_computation}. MZMs usually arise as ($d-1$)-dimensional edge modes of $d$-dimensional topological superconductors\cite{TSC_edge1,TSC_edge2,TSC_edge3,TSC_edge4} (TSCs), and the classification for TSCs has been thoroughly analyzed for different symmetry classes in different dimensions\cite{TSC_classification1,TSC_classification2}. In recent years, many theories have been proposed to realize MZMs\cite{TSC_Kitaev,TSC_Fu,CuBiSe_Fu,Chiral_Zhang,TSC_Das,TSC_Oppen,TSC_Fisher,Sau2010,Huge_vortex1,Huge_vortex2,magchain_Bernevig,TSC_Nagaosa,TSC_Wu,TSC_XuG,skymion,TSC_Liu1,TSC_Liu2,TSC_Wang} and signatures for the existence of MZMs in superconductors (SCs) have also been observed in experiments\cite{TSC_Ando,TSC_Jia,TSC_Mourik,TSC_Shtrik,TSC_Xu,TSC_Furdyna,magchain_exp,chiral_exp,TSC_iron1,TSC_iron2}.

1D $p$-wave spinless SCs and 2D $p+ip$-wave SCs are examples of TSCs\cite{quantum_computation,TSC_edge1,TSC_edge2,TSC_edge3,TSC_edge4,TSC_Kitaev}. However, the existence of $p$-wave or $p+ip$-wave SCs has not been experimentally confirmed without a doubt. The discovery of topological insulators\cite{2DTI_Kane1,2DTI_Kane2,2DTI_Kane3,2DTI_Zhang} (TIs) is a breakthrough in the pursuit for TSCs. It is found that, the topologically protected surface Dirac cone of a 3D strong topological insulator (STI) combined with $s$-wave superconductivity, is equivalent to a $p+ip$-wave spinless SC\cite{TSC_Fu}. As a result, a single MZM can be trapped in a vortex in such a system. Soon after, \emph{P. Hosur et al.} introduce the concept of vortex phase transitions (VPTs), and point out that if a STI can be doped to be bulk superconducting, there can also be a single MZM at the vortex core on its surface\cite{Vishwanath_vortex}. Based on the concept of VPTs, a recent work points out that a 1D robust gapless phase with pairs of bulk MZMs, can be realized in doped superconducting topological Dirac semimetals (TDSs)\cite{TDS_vortex}. Besides the well-known candidates for the STIs and TDSs\cite{TSM_rev,Dirac Kane,Na3Bi,Cr3As2}, there is another interesting class of topological systems, the weak topological insulators (WTIs)\cite{TSC_edge4,3DTI}, which can be viewed as 2D TIs sticked together weakly in the third direction. However, there are very few candidates for 3D WTIs. The study of the VPTs in a superconducting 3D WTI has not been carried out.

Recently, the topological properties in iron-based  SCs predicted theoretically\cite{iron_Hao,FeTeSe_3D,FeTeSe_monolayer} have gained a major support by new experimental evidence\cite{TSC_iron1,TSC_iron2} in FeSe$_{1-x}$Te$_x$. It is possible that iron-based SCs can be unique promising systems to realize MZM at much higher temperature. The MZMs arise in the vortices in FeSe$_{1-x}$Te$_x$ because there is an intrinsic  band inversion at the Z point. The band inverson leads to a 3D STI state with topologically protected surface states. Following the Fu-Kane proposal\cite{TSC_Fu}, the vortex in the superconducting surface states naturally hosts a MZM phase. However, iron-based superconductors are known to be versatile for their structures. In particular, the electronic structure and layer coupling along $c$-axis can be varied significantly. For example, a zero bias peak is also observed in another iron-based SC, (Li$_{0.84}$Fe$_{0.16}$)OHFeSe\cite{FeSe_feng}, whose electronic structure  differs significally from FeSe$_{1-x}$Te$_x$ in  three major ways: (1) the former is with much heavier electron doping than the latter; (2) the former has no hole pocket at the $\Gamma$ point; (3) the coupling between FeSe layers is much weaker in the former than in the latter. These differnces question  whether (Li$_{0.84}$Fe$_{0.16}$)OHFeSe can be treated as a 3D STI.

In this paper, we show that there are two additional VPTs in iron-based SCs as superconducting 3D WTI. First, for  the VPTs in a superconducting 3D WTI with both time reversal symmetry (TRS) and inversion symmetry (IS), we derive the following essential results. Depending on crystalline symmetries and orbitals contributing to the band inversions, the superconducting 3D WTI with a single quantum vortex line can be classified into two cases: (1) if the band inversions occur on a $C_n$ ($n>2$) rotational axis and  the angular momenta of the orbitals contributing to the band inversions are not equivalent, the superconducting vortex line has a robust nodal phase with pairs of bulk MZMs; otherwise (2) the vortex line is fully gapped and there exist vortex end MZMs in a certain range of doping level. In both cases, the topological phase transition can take place by tuning chemical potentials or Fermi levels. The range of the chemical potential corresponding to the topologically nontrivial phase is proportional to the band dispersion along the weak coupling direction. We also extend our conclusions to a class of normal insulators (NIs). Second, we show that the above theory can be applied to classify the topological vortex phases and understand recent experimental results on iron-based SCs.  Both band inversions at the $\Gamma$ and M points in the band strucure of iron-based SCs can generate topological VPTs. For example, the nodal superconducting vortex line phase can be realized by further electron doping in FeSe/Te  and LiFeAs \cite{iron_TDS}. For (Li$_{1-x}$Fe$_{x}$)OHFeSe, different topological vortex phases can be realized by changing $x$.   It is also possible that the vortex end MZMs on the (001) surface (Li$_{0.84}$Fe$_{0.16}$)OHFeSe can stem from the 3DWTI state caused by the band inversion at M points discussed in Ref.\cite{iron_Hao}.

\section {\bf General theory for 3D WTIs}

We consider  a general spinful system with intrinsic $s$-wave spin-singlet superconductivity.  The superconducting Hamiltonian in the basis $\Psi^\dag({\bf k})=(c_{\uparrow}^\dag({\bf k}),c_{\downarrow}^\dag({\bf k}),c_{\uparrow}(-{\bf k}),c_{\downarrow}(-{\bf k}))$ can be written as
\begin{eqnarray}\label{Hsc}
H_{sc}&=&\left(\begin{array}{cc}
          H_0(\mathbf{k})-\mu     &     \Delta                        \\
          \Delta^\dagger          &     \mu-H_0^\ast(-\mathbf{k})     \\
          \end{array}
          \right),
\end{eqnarray}
where we have only preserved the indices for the spin and Nambu spaces. Here, $\Delta$ is the superconducting order parameter and takes the form $\Delta_0i\sigma_2$ in the spin space, $\mu$  is the chemical potential, and $H_0(\mathbf{k})$  is the normal state Hamiltonian. If a single quantum vortex goes through the system along the $z$-direction, the superconducting order parameter $\Delta(\mathbf{r})$ in the real space is  transformed into $\Delta(r)e^{i\theta}$, where $r=\sqrt{x^2+y^2}$ and $\theta$ is the polar angle\cite{Vishwanath_vortex,Fang_vortex}. Obviously, the vortex line destroys the translational symmetry in the $xy$-plane. The TRS is broken while the particle-hole symmetry (PHS) is preserved. Consequently, such a spinful superconducting system with a single quantum vortex line belongs to the class $D$ of the Altland-Zirnbauer classification\cite{TSC_classification1,TSC_classification2}.

We first analyze the topological phase of such a superconducting vortex line by assuming that it is a full-gap system. For a full-gap class $D$ system, it has a $Z_2$ topological classification.  The topological invariance is defined as
\begin{eqnarray}\label{Z2}
\nu=sgn(Pf(H_{sc}(k_z=0))\cdot Pf(H_{sc}(k_z=\pi))),
\end{eqnarray}
where $sgn()$ is the signum and $Pf()$ is the Pfaffian of a antisymmetric matrix\cite{TSC_classification1,TSC_classification2,Topo_D}. In the following, we take $Pf(k_z)$ for $Pf(H_{sc}(k_z))$ for simplicity. From Eq.\ref{Z2},  the topological VPTs of such a system are merely determined by the gap close-and-reopen processes at $k_z=0, \pi$. Moreover, it has been concluded that \cite{TDS_vortex} if the normal state Hamiltonian describes a 2D insulator in the $k_z=0/\pi$ plane, $sgn(Pf(0))/sgn(Pf(\pi))$ is simply given by the $Z_2$ topological invariance of the 2D insulator in the $k_z=0/\pi$ plane, and if it is a TI in the $k_z=0/\pi$ plane, $sgn(Pf(0))/sgn(Pf(\pi))$ must change sign at some critical chemical potential $\mu_c(0)/\mu_c(\pi)$.

For a superconducting WTI, if the vortex line is along the weak coupling direction, assuming that the system is full-gap, we can conclude that the superconducting vortex line is topologically trivial if the chemical potential is in the insulating gap of the normal state, and that the vortex line is nontrivial if the chemical potential is between $\mu_c(0)$ and $\mu_c(\pi)$. In the latter case, there are vortex end MZMs even though there exist no topologically nontrivial surface states.

To be specific, we consider a lattice model
\begin{eqnarray}\label{H0_WTI}
H_0^{(1)}&=&(m-t\cos k_x-t\cos k_y-t_3\cos k_z)\Sigma_{30} \nonumber\\
&+&t^\prime \sin k_x\Sigma_{11}+t^\prime \sin k_y\Sigma_{12}+t_3^\prime \sin k_z\Sigma_{13},
\end{eqnarray}
in which $\Sigma_{ij}=\tau_i\sigma_j$, and the parameters are taken as $\{t, t_3, t^\prime, t_3^\prime\}=\{1.0, 0.5, 1.0, 1.0\}$. Here, $\tau_i$ and $\sigma_i$ ($i=0, 1, 2, 3$) are Pauli matrices representing the orbital and spin degrees of freedom. This model respects the TRS and the full symmetry of the $D_{4h}$ point group. The related symmetry group
generators are given by the following matrices: $I=\Sigma_{30}$, $C_{4z}=\tau_0\otimes e^{i\sigma_z\pi/4}$, $C_{2x}=\tau_0\otimes e^{i\sigma_x\pi/2}$ and $T=i\Sigma_{02}K$, where $K$ is the complex conjugate operation. Obviously, the basis of the above model contains two Kramers' doublets with opposite parity, and each Kramers' doublet have angular momentum $\pm\frac{1}{2}$. In the calculations, we set $m=1.2$ to derive a WTI. With these parameters, $H_0^{(1)}$ has two band inversions at the $\Gamma$ point and Z point, and it has no topologically nontrivial surface states on the $(001)$ surface while there are two Dirac cones on the $(100)$ surface.

%
%

\begin{figure}
\centerline{\includegraphics[width=0.45\textwidth]{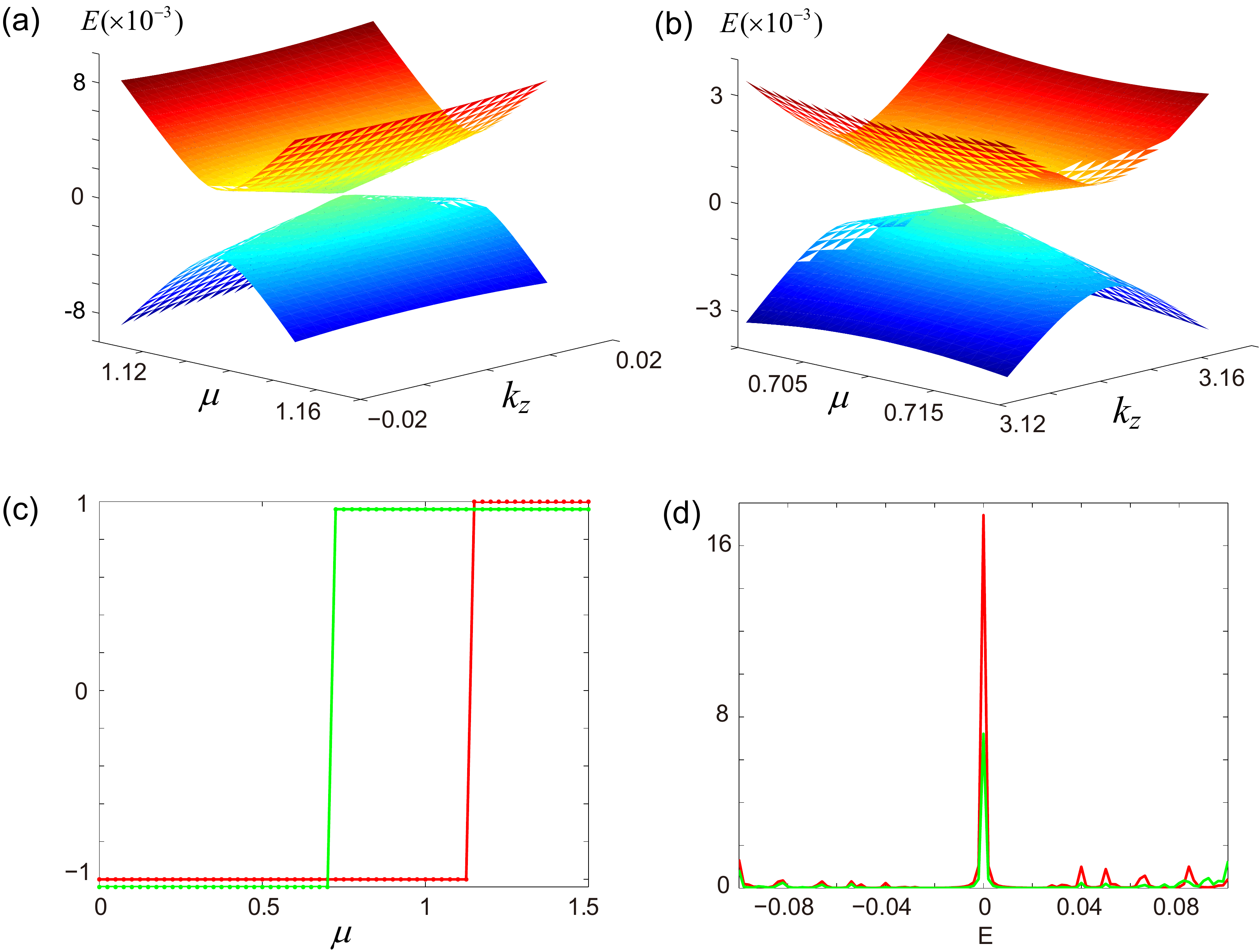}}
\caption{(color online) Results of a quantum vortex line in the superconducting state of the WTI in $H_0^{(1)}$. (a) and (b): lowest energy $E$ of the
vortex line as a function of the chemical potential $\mu$ near $k_z=0$ and $k_z=\pi$, respectively (the lattice size in the $xy$-plane is $24\times 24$). Note that $E=0$ occurs at some critical value of $\mu$ for $k_z=0$ and $\pi$. (c): $sgn(Pf(k_z=0))$ (red) and $sgn(Pf(k_z=\pi))$ (green) as functions of $\mu$. (the lattice size in the $xy$-plane is $16\times 16$). The green line is shifted down by $0.04$ for visual purpose. (d) the spin resolved LDOS at $\mu=1.0$, red for spin-up and green for spin-down, at the vortex core on the $(001)$ surface. (the lattice size is $14\times14\times26$). In the calculations, the parameters are $\{t, t_3, t^\prime, t_3^\prime\}=\{1.0, 0.5, 1.0, 1.0\}$, $m=1.2$, $\Delta_0=0.2$, and the size of the vortex $R\rightarrow 0$.
\label{vertex_WTI}}
\end{figure}

Assuming on-site intra-orbital pairing, we analyze the topological VPTs in the WTI depicted by $H_0^{(1)}$ numerically. In the calculations, we take a simple form of $\Delta(r)$: $\Delta(r)=\Delta_0\Theta(r-R)$, where $\Theta(r)$ is the step function and $R$ is the vortex size. By diagonalizing the Hamiltonian, we obtain the energy spectrum in the superconducting vortex line. It turns out that the vortex line is indeed fully gapped. If we tune the chemical potential continuously, the energy spectrum of the system becomes gapless at $\mu^c_1=1.14$ ($\mu^c_2=0.716$) at $k_z=0$ ($k_z=\pi$), shown in Fig.\ref{vertex_WTI}(a) and (b), indicating that the system goes through two topological VPTs. By calculating the $Z_2$ topological index directly, we find that a topological phase transition occurs when $\mu^c_1=1.15$ ($\mu^c_2=0.7$) at $k_z=0$ ($k_z=\pi$), and the system is topologically nontrivial when $\mu^c_2<\mu<\mu^c_1$, as shown in Fig.\ref{vertex_WTI}(c). The results of the two different methods are consistent. In the topologically nontrivial phase, there must be vortex end MZMs on the $(001)$ surface.  To show this, we calculate the local density of states (LDOS) near the vortex core on the $(001)$ surface
\begin{eqnarray}\label{LDOS}
\rho_\sigma({\bf r_0},E)&=&\sum_{a,n}\int d{\bf r}\langle\varphi_n({\bf r})|c_{a,\sigma}^\dag({\bf r}) c_{a,\sigma}({\bf r})|\varphi_n({\bf r})\rangle \nonumber\\
&&\delta({\bf r}-{\bf r_0})\delta(E_n-E),
\end{eqnarray}
where $a$ and $\sigma$ label orbital and spin respectively, $n$ is the quasi particle spectrum index. Fig.\ref{vertex_WTI}(d) shows the result for $\mu=1.0$. Obviously, there is a zero-bias peak (ZBP) at the vortex core. Moreover, the ZBP is contributed by different spin components unequally, which leads to spin polarized zero-bias conductance peak in scanning tunneling microscope (STM) experiment\cite{STM1,STM2,STM3}.

It is worth to mention that, the topological VPTs in the superconducting WTI can also be determined by the Berry phase framework in Ref.\cite{Vishwanath_vortex}: the VPT point corresponds to such a condition that the eigenvalues of the $SU(2)$ Berry connection on the Fermi surfaces  in the $k_z=0/\pi$ plane, are $\pm\pi$. Based on this criterion, we have roughly estimated the topological phase transitions by expanding  the Hamiltonian in Eq.\ref{H0_WTI} in the continuum limit. As expected, there are two VPT points: $\mu^c_1=t^\prime\sqrt{2(2t-m+t_3)/t}$ at $k_z=0$ and $\mu^c_2=t^\prime\sqrt{2(2t-m-t_3)/t}$ at $k_z=\pi$. Apparently, the range of the topologically nontrivial phase is proportional to $t_3$, namely the coupling strength along the weak coupling direction. It should be emphasized that, if the condition of the superconductivity and the Fermi surfaces are complicated, the Berry phase criterion in Ref.\cite{Vishwanath_vortex} fails to determine the VPT points and we need to calculate the energy spectrum or the $Z_2$ topological invariant of the vortex line to obtain the VPTs\cite{Huge_vortex2,TSC_XuG,TDS_vortex}.

From the above analysis, we can achieve topologically nontrivial full-gap vortex line phase in the superconducting WTIs.  However, do all the WTIs have similar results? The answer is negative. In fact, the system may not be full-gap\cite{TDS_vortex}.  In the following, we show that there is a class of WTIs whose superconducting state with a single quantum vortex line has a robust nodal superconductor phase, rather than a topologically nontrivial full-gap phase. To be specific, we consider the following lattice model
\begin{eqnarray}\label{H0_WTI2}
H_0^{(2)}&=&(m-t\cos k_x-t\cos k_y-t_3\cos k_z)\Sigma_{30}  \nonumber\\
&+&t^\prime \sin k_x\Sigma_{13}+t_3^\prime \sin k_z(\cos k_x-\cos k_y)\Sigma_{11}  \nonumber\\
&-&t^\prime \sin k_y\Sigma_{20}+2t_3^\prime \sin k_z\sin k_x\sin k_y\Sigma_{12}.
\end{eqnarray}
Here, $H_0^{(2)}$ respects the same symmetries as $H_0^{(1)}$ in Eq.\ref{H0_WTI}, and the only difference between them is that, one of the Kramers' doublets in the basis is changed into one  with the same parity but a different  angular momentum quantum number, $\pm\frac{3}{2}$. Correspondingly, the matrix form of the symmetry group generators transforms into $C_{4z}=(\Sigma_{30}+i\Sigma_{03})/\sqrt{2}$ and $C_{2x}=\Sigma_{31}$, while both the IS and TRS remain unchanged. We take all the parameters in $H_0^{(2)}$ the same as those in $H_0^{(1)}$, which leads to that the two Hamiltonians have similar topological property.



\begin{figure}
\centerline{\includegraphics[width=0.45\textwidth]{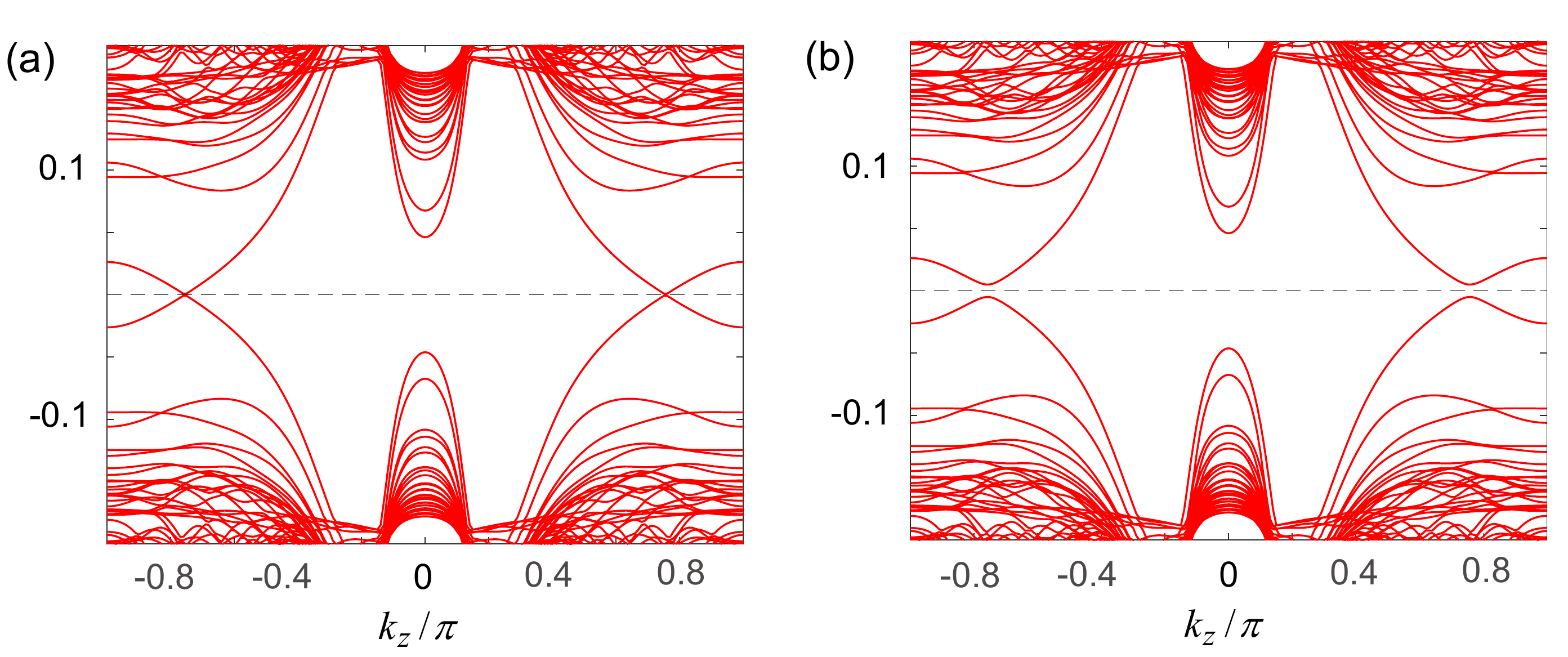}}
\caption{(color online) Dispersion of the low-energy bands of the vortex line in a superconducting WTI as a function of $k_z$ for the chemical potential $\mu=1.0$. (a) The WTI is depicted by $H_0^{(2)}$. Note that there are two nodes at $k_z=\pm0.75\pi$. (b) The WTI is described by $H_0^{(2)}$ with an additional $C_{4z}$ rotational symmetry breaking term $t_{sb}\sin k_z\Sigma_{11}$, where $t_{sb}=0.1$. Obviously, the vortex line becomes fully gapped. In the calculations, all other parameters are the same as in Fig.\ref{vertex_WTI}.
\label{TWSC}}
\end{figure}

However, as we shall show below, the topological property of the vortex line in the superconducting state of $H_0^{(2)}$ is completely different from that in $H_0^{(1)}$. In addition to the topological VPTs at $k_z=0$ and $k_z=\pi$, which occur at exactly the same critical chemical potential $\mu^c_1$ and $\mu^c_2$ as in the case of $H_0^{(1)}$, for any given $k_z$, there is also a chemical potential $\mu$ so that the vortex line in the superconducting WTIs described by $H_0^{(2)}$ becomes gapless. In Fig.\ref{TWSC}(a), we show this explicitly for $\mu=1.0$. Obviously, the vortex line has two nodes at $\pm0.75\pi$. Moreover, the nodes in the vortex line are rather stable: the nodal points can never be gapped out unless two points meet with each other. In the limit $t_3^\prime=0$, we can get the trajectory of the nodal points in the $(k_z, \mu)$ space. We begin with an infinite large chemical potential $\mu$. As it decreases, a nodal point first emerges at $k_z=0$ when $\mu=\mu^c_1$. Then the nodal point splits into two and they move in the opposite directions on the $k_z$-axis. Finally the two nodal points meet with each other at $k_z=\pi$ when $\mu=\mu^c_2$ and gap out. A finite $t_3^\prime$ only changes the condition quantitatively where the nodal points emerge and vanish. Therefore, this is a robust 1D nodal superconducting phase with bulk MZMs\cite{TDS_vortex}. Actually, this difference stems from the fact that, there is a band inversion at every $k_z$ on the $\Gamma$-Z line for $H_0^{(2)}$ while it is true only at the $\Gamma$ point and Z point for $H_0^{(1)}$. If a $C_{4z}$ rotational symmetry broken perturbation $t_{sb}\sin k_z\Sigma_{11}$ is added, $H_0^{(2)}$ still describes a WTI with band inversions at the $\Gamma$ point and Z point. However, the band inversion on the $\Gamma$-$Z$ line ($k_z\neq 0, \pi$) is no longer true. Namely the system is similar to that in $H_0^{(1)}$. Consequently, the energy spectrum of the corresponding vortex line system becomes fully gapped, shown in Fig.\ref{TWSC}(b).


From the above results, we can find that the WTIs can be classified into two classes according to the topological property of their superconducting vortex line. The WTIs described by $H_0^{(1)}$ and $H_0^{(2)}$ can be viewed as the generalization of the 3D STIs and TDSs respectively. To be specific, the two Hamiltonians both describe 2D TIs if we omit the coupling in the $z$-direction. With the coupling terms in the $z$-direction turned on, especially the $t_3$ term, both of the two Hamiltonians describe 3D WTIs with band inversions at the $\Gamma$ point and $Z$ point when $t_3$ is small. As $t_3$ becomes larger, the WTIs have larger band dispersion in the $k_z$-direction and the insulating gap closes at the Z point at some critical $t_3=t^c$. Finally, the WTIs in $H_0^{(1)}$ and $H_0^{(2)}$ evolve into STIs and TDSs with a band inversion at the $\Gamma$ point respectively with $t_3$ tuned to be larger. As a result, it can be inferred that the superconducting vortex line for the WTIs in $H_0^{(1)}$ and $H_0^{(2)}$ must have similar topological properties with the STI\cite{Vishwanath_vortex} and TDS\cite{TDS_vortex} case respectively. Recalling the condition for the existence of TDSs\cite{DS_Nagosa}, we can straightforwardly conclude that the nodal superconducting vortex line phase is stable only when the band inversions of the WTI occur between two bands with nonequivalent angular momentum quantum numbers defined by the $C_n$ ($n>2$) rotational axis, and in the other case the superconducting WTI plus a single quantum vortex line is full-gap with a single vortex end MZM for certain range of doping level. Based on the above results, we can summarize the topological VPTs for the STIs, TDSs and WTIs in a general phase diagram shown in Fig.\ref{phase}(a).


\begin{figure}
\centerline{\includegraphics[width=0.45\textwidth]{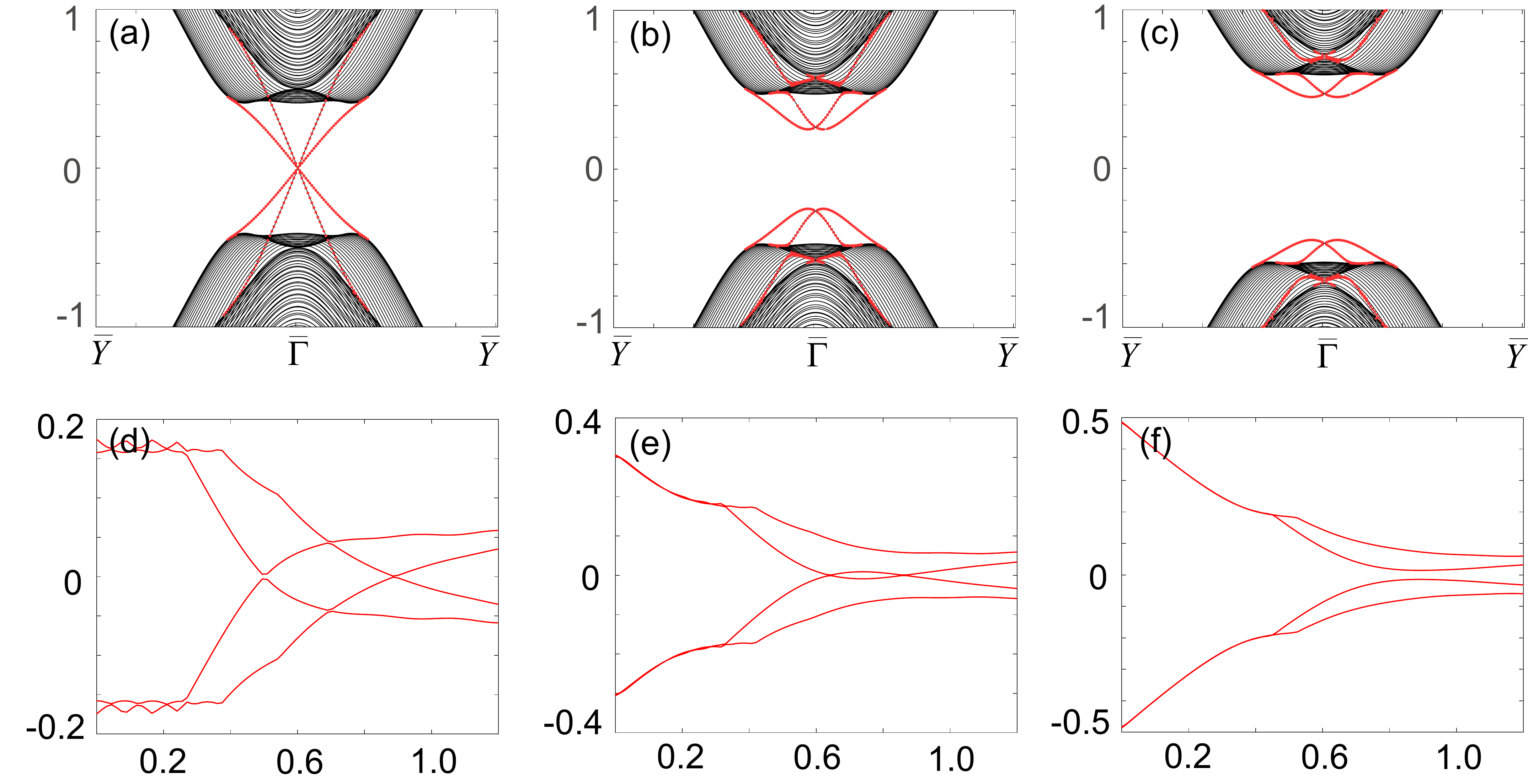}}
\caption{(color online) The edge states along the high-symmetry line on the (100) edge for $H_0^{DB}$ are presented in (a)$\sim$(c). (d)$\sim$(f) show the lowest energies of the 0D superconducting vortex system corresponding to $H_0^{DB}$. (a)(d) show the results for the $\lambda=0$ case, (b)(e) for the $\lambda=0.14$ case, and (c)(f) for $\lambda=0.25$. In the calculations, the parameters for the superconductivity are the same as in Fig.\ref{vertex_WTI} with the lattice size  $20\times 20$.
\label{WTI2}}
\end{figure}

\section {\bf generalization to $Z_2$ trivial insulators with double-band inversions}

Based on the above analysis and our previous work in Ref.\cite{TDS_vortex}, the topologically nontrivial vortex line state is closely related to the topological property of the normal state, which is determined by the band inversions in the Brillouin zone (BZ). In the 3DWTI case, there are two band inversions which occur at different $k$-points in the BZ. However, there is another possibility that the two band inversions occur at the same $k$-point in the BZ, the so-called double-band inversion. In the normal state, such a double-band inversion can not change the $Z_2$ topological invariant of the system. However, it can lead to some topologically nontrivial crystalline state, such as the topological mirror insulator state in the antiperovskites A$_3$SnO \cite{TCI_antiperovskite1,TCI_antiperovskite2} and wallpaper fermion state in Sr$_2$Pb$_3$\cite{wallpaper}. Therefore, it is natural to ask whether the topologically nontrivial vortex line state can be realized in systems with double-band inversion. To illustrate this, we introduce the following model
\begin{eqnarray}\label{H0_WTI_2D}
H_0^{DB}&=&\left(\begin{array}{cc}
          H_{01}           &     H_{hy}     \\
          H_{hy}^\dagger   &     H_{02}     \\
          \end{array}
          \right), \\
H_{01}&=&m_1(k)\Sigma_{30}+\nu_1[\sin k_x\Sigma_{13}+\sin k_y\Sigma_{20}], \nonumber\\
H_{02}&=&m_2(k)\Sigma_{30}+\nu_2[\sin k_x\Sigma_{20}-\sin k_y\Sigma_{13}], \nonumber\\
H_{hy}&=&\lambda(i\sigma_x\tau_++i\sigma_y\tau_-), \nonumber
\end{eqnarray}
where $m_i(k)=M_i-B_i(\cos k_x+\cos k_y)$ ($i=1, 2$), $\tau_\pm=\tau_x\pm i\tau_y$, and the parameters are set to be $\{\nu_1, \nu_2, M_1, M_2, B_1, B_2\}=\{0.5, 1.0, 1.5, 1.5, 1.0, 1.0\}$. The above Hamiltonian can be regarded as two copies of 2D TIs when $\lambda=0$, and $H_{hy}$ is the hybridization between the two TIs respecting both the TRS and IS. The TRS and IS take the matrix form $T=Ki\Sigma_{02}\gamma_0$ and $I=\Sigma_{30}\gamma_3$ respectively, where $\gamma_i$ ($i=1, 2, 3$) are the Pauli matrices defined in the space constructed by $H_{01}$ and $H_{02}$. Obviously, when $\lambda=0$, there will be two Dirac cones on the edge, as shown in Fig.\ref{WTI2}(a). When $H_{hyb}$ is turned on, the two TIs hybridize  and the Dirac cones on the edge hybridize as well to open a gap, shown in Fig.\ref{WTI2}(b)(c). Therefore, $H_0^{DB}$ remains topologically trivial for any $\lambda$ value.

Considering  an on-site intra-orbital pairing, we analyze its topological VPTs. As mentioned above, $H_0^{DB}$ can be viewed as two independent 2D TIs when  $\lambda$ is zero.  Each of the two superconducting 2D TIs plus a single quantum vortex must go through a VPT at some critical chemical potential $\mu^c$\cite{TDS_vortex}. As a result, if the two TIs have different $\mu^c$, for example $\mu^c_1$ ($\mu^c_2$) for $H_{01}$ ($H_{02}$), the superconducting vortex becomes topologically nontrivial when the chemical potential is between $\mu^c_1$ and $\mu^c_2$. With the parameters above, $\mu^c_1$ and $\mu^c_2$ are determined to be $0.50$ and $0.89$, shown in Fig.\ref{WTI2}(d). When a small $\lambda$ is turned on, the two TIs hybridize with each other. However, the topologically nontrivial vortex state can not be destroyed immediately as the topological phase is protected by a finite energy gap. As shown in Fig.\ref{WTI2}(e), there are still two VPT points at $\mu^c_1=0.63$ and $\mu^c_2=0.87$ when $\lambda=0.14$.  The topologically nontrivial vortex line phase  becomes smaller when the hybridization becomes strong. Eventurally, it becomes topologically trivial if the $\lambda$ is strong enough, as shown in Fig.\ref{WTI2}(f).

\begin{figure}
\centerline{\includegraphics[width=0.45\textwidth]{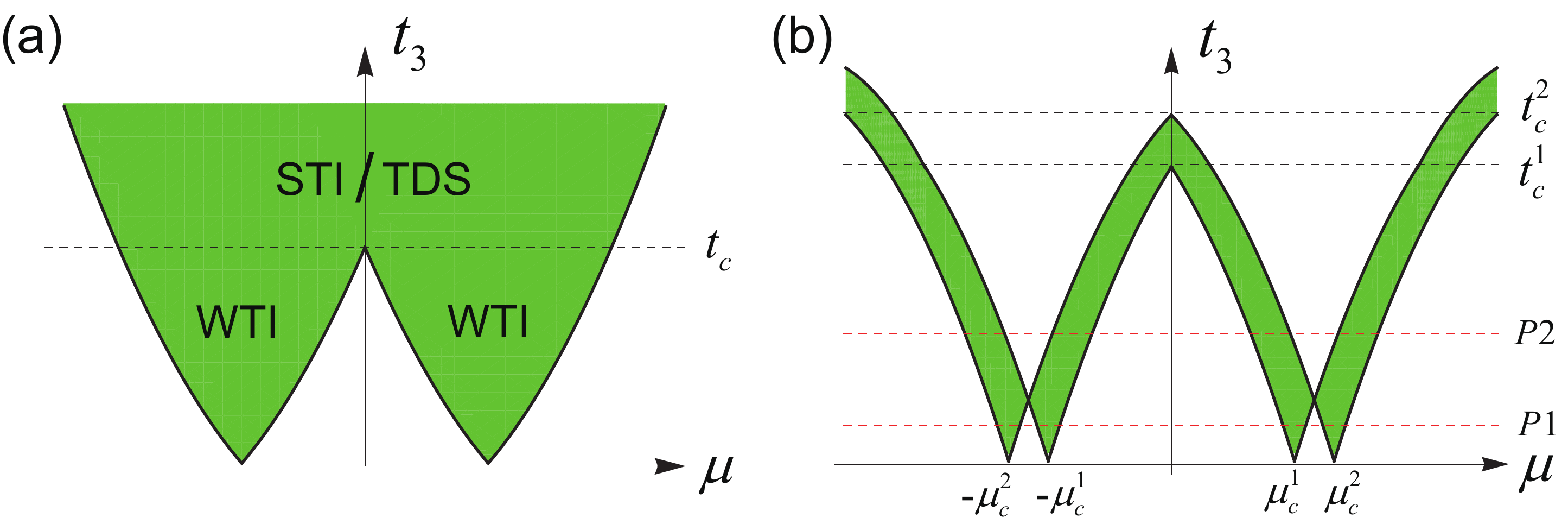}}
\caption{(color online) Illustrative topological phase diagrams of the superconducting vortex line as a function of $t_3$ and the chemical potential $\mu$ for the 3D WTIs in Eq.\ref{H0_WTI} and Eq.\ref{H0_WTI2} in (a) and for the NIs with double-band inversion in Eq.\ref{H0_WTI_2D} in (b). The green region represents the full-gap topologically nontrivial vortex line phase or the quasi-1D nodal superconducting phase, depending on the basis and the symmetry group of the system.
\label{phase}}
\end{figure}

Take the coupling in the $z$-direction into consideration, for instance adding $t_3\cos k_z$ into $m_i(k)$, we can achieve a topological phase diagram for the superconducting vortex line for the systems with double-band inversions similar to that in $H_0^{DB}$, shown in Fig.\ref{phase}(b). As analyzed above, there are two VPTs at $\mu^c_1$ and $\mu^c_2$ at $k_z=0$ resulting from the double-band inversion at the $\Gamma$ point. In the limit $t_3=0$, a same double-band inversion occurs at the Z point, leading to two VPTs at $\mu^c_3=\mu^c_1$ and $\mu^c_4=\mu^c_2$ at $k_z=\pi$. The superconducting vortex line is always topologically trivial. As a small $t_3$ is turned on, the VPTs at $k_z=\pi$ and $k_z=0$ take place at different chemical potential values.  Consequently, there will be two narrow topologically nontrivial phases when $\mu_3^c<\mu<\mu_1^c$ and $\mu_4^c<\mu<\mu_2^c$ in the electron-dope region, as the $P1$ phase shown in Fig.\ref{phase}(b). The region of the phase  is proportional to strength of $t_3$. If $t_3$ is tuned to be larger, $\mu^c_4$ becomes smaller than $\mu^c_1$ and the superconducting vortex line will be topologically nontrivial when $\mu_3^c<\mu<\mu_4^c$ and $\mu_1^c<\mu<\mu_2^c$, which is the $P2$ phase shown in Fig.\ref{phase}(b). When $t_3$ is large enough ($t_3>t_c^2$), the double-band inversion at the Z point vanishes, namely there is no VPT at $k_z=\pi$ and the superconducting vortex line is topologically trivial only when $\mu_1^c<\mu<\mu_2^c$. Remarkably, the above topologically nontrivial superconducting vortex line is full-gap when and only when $H_{01}$ and $H_{02}$ in $H_0^{DB}$ both describe STIs or WTIs similar to that in $H_0^{(1)}$, and there can be vortex end MZMs for a certain range of doping level in this case. If at least one of the $H_{01}$ and $H_{02}$ describe WTIs similar to that in $H_0^{(2)}$, the system has a 1D robust nodal superconducting phase.


\begin{figure}
\centerline{\includegraphics[width=0.45\textwidth]{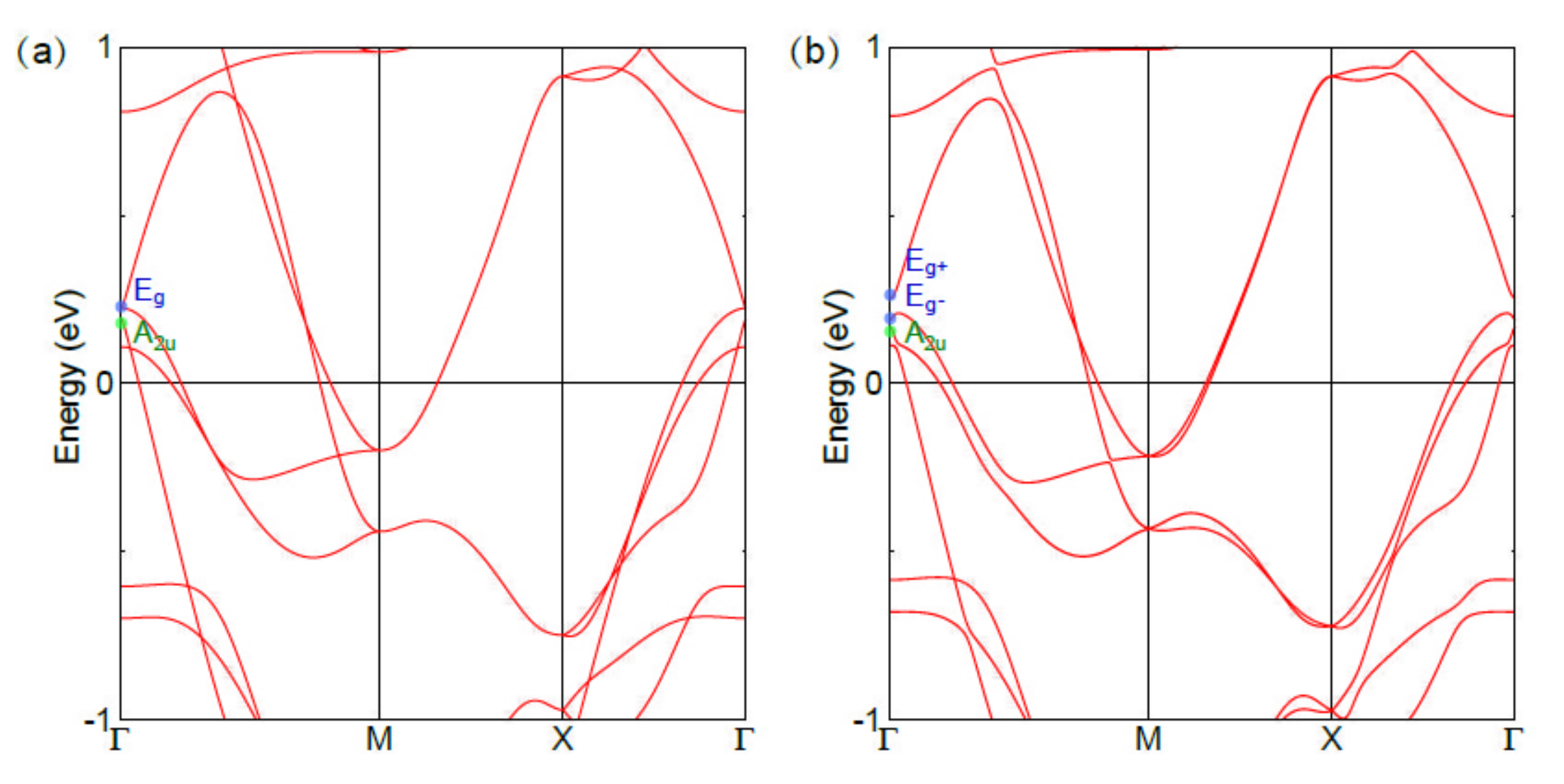}}
\caption{(color online) The band structures for LiOHFeSe without SOC (a) and with SOC (b). The bands at $\Gamma$ point which belong to the $E_g$ ($A_{2u}$) irreducible representation are marked by the blue (green) point.
\label{iron_band}}
\end{figure}

\section {\bf Topological VPTs in iron-based SCs}

Now we apply the above theory to iron-based SCs. We first focus on the band inversion at the zone center. We take LiOHFeSe as an example, and calculate its band structure with the experimental lattice parameters for (Li$_{0.8}$Fe$_{0.2}$)OHFeSe\cite{lattice_FeSe}, which is shown in Fig.\ref{iron_band}. Without spin-orbital coupling (SOC), there are three major bands near the $\Gamma$ point, which are labeled as the $A_{2u}$ band and the two-fold degenerate $E_g$ bands, shown in Fig.\ref{iron_band}(a). With SOC, the $E_g$ bands split into a $E_{g+}$ band and a $E_{g-}$ band, as shown in Fig.\ref{iron_band}(b). The band inversions between the $A_{2u}$ and $E_{g\pm}$ bands can create  nontrivial topological properties as they have opposite parities\cite{Z2 inversion}. In the tetragonal lattice structure in which the $C_4$ symmetry is preserved,   the angular momentum quantum numbers for  these bands are $\pm3/2$ for $E_{g+}$, $\pm1/2$ for $E_{g-}$ and $\pm1/2$ for  $A_{2u}$. Thus, we can directly apply the above theory to obtain the topological vortex phases caused by the  band inversion at $\Gamma$ point.

\begin{figure}
\centerline{\includegraphics[width=0.45\textwidth]{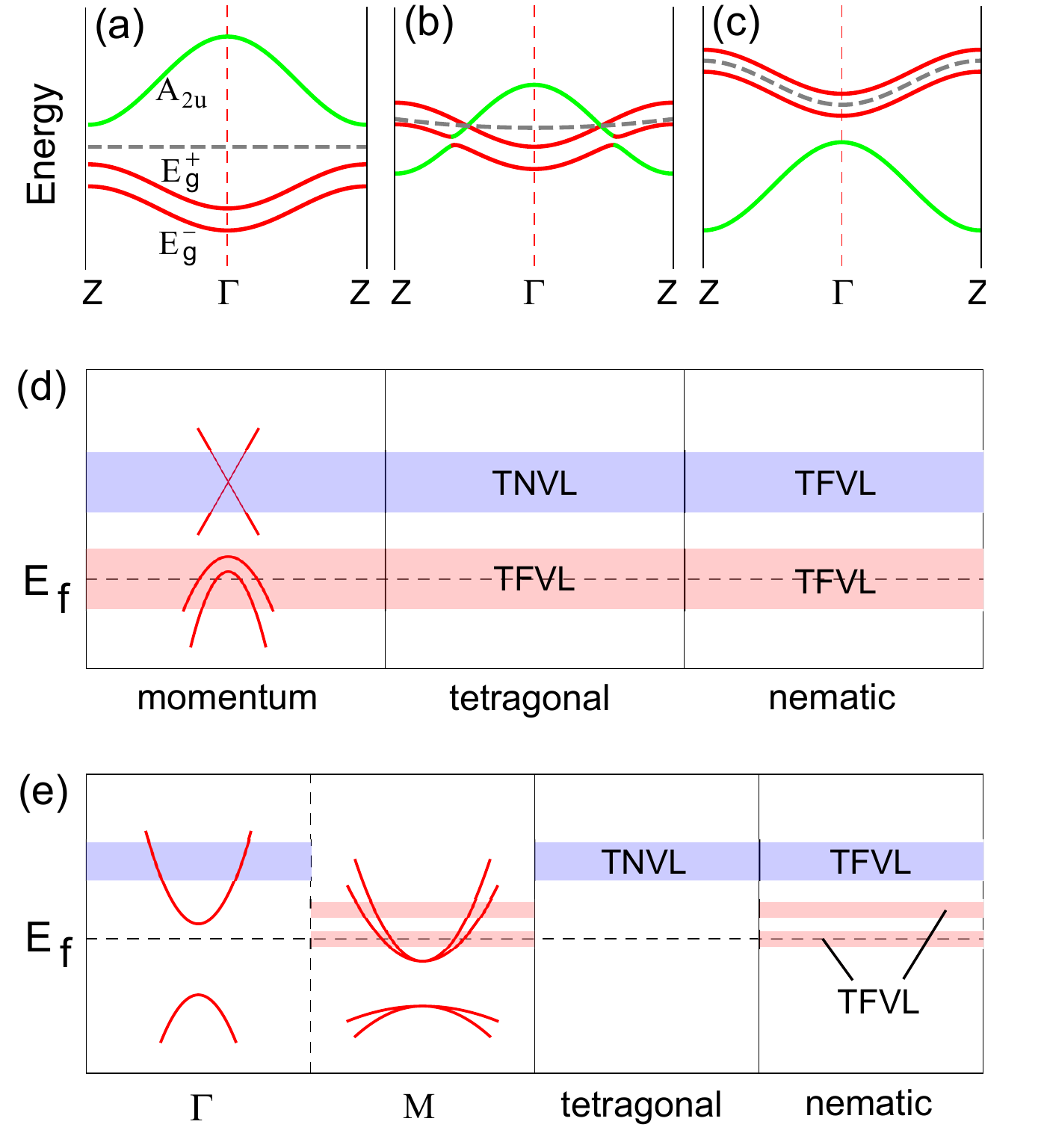}}
\caption{(color online) Sketch of the band structures at Brillouin Zone center for  typical iron-based superconductors and topological phases. (a) Topological trivial band structure; (b) Strong TI or Dirac Semimetal phases; (c) 3D Weak TI case; (d) Topological vortex phases for the (b) case; (e) Topological vortex phases for the (c) case.  The case corresponding to the double-band inversion at the BZ corner M point is also present in (e).
\label{iron_phase}}
\end{figure}

The above electronic band structure is qualitively common acrossing all families of iron-based SCs. There are only quantitative differences between different SCs.  The major quantiative parameters that can affect the topological properties include the interlayer coupling along $c$-axis, the relative on-site energy difference between the $A_{2u}$ and $E_g$ bands,  the SOC strength and the chemical potential (Fermi energy or carrier density). Depending on these four  energy scales, we sketch three typical band structures in Fig.\ref{iron_phase}(a)(b)(c) for iron-based SCs.  Fig.\ref{iron_phase}(a) represents a trival phase without a band inversion at high symmetry points. In Fig.\ref{iron_phase}(b), the band inversion between $A_{2u}$ and $E_g$ bands takes place at Z point but not at $\Gamma$ point. In this case, it represents a 3D STI phase or a TDS phase depending on the chemical potential.  If the Fermi energy is set in between the $E_{g-}$ and $A_{2u}$ $(E_{g+})$ bands at the Z point, it is STI (TDS). The topological vortex phases in this region are sketched in Fig.\ref{iron_phase}(d). There are two phases, the topological nodal vortex line (TNVL) phase and topological full-gap vortex line (TFVL) phase with vortex end MZMs, which result from TDS and STI states respectively.  A nematic order, which breaks the $C_4$ rotation symmetry, can drive VPT from the TNVL phase to TFVL phase as shown in Fig.\ref{iron_phase}(d).

In Fig.\ref{iron_phase}(c), the band inversion between the $A_{2u}$ and $E_g$ bands takes place at both $\Gamma$ and Z points, which is corresponding to a 3DWTI phase. In this case, the band structure and the topological vortex line phases are sketched in Fig.\ref{iron_phase}(e).  In fact, the band of LiOHFeSe shown in Fig.\ref{iron_band} belongs  to this case, in which the $E_{g+}$ band and the $A_{2u}$ band has a band inversion at both $\Gamma$ and Z point.  The band inversion is mainly caused by the strong coupling between the Fe-$d_{xy}$ orbital and the Se-$p_z$ orbital in the FeSe layer. The band inversion is known to takes place much easylier in a material with a smaller in-plane lattice parameter. As pointed out in Ref.\cite{FeTeSe_monolayer}, the band inversion occurs when the in-plane lattice parameter is smaller than $3.905$ ${\AA}$. This is consistent with the Li(OH)FeSe case, whose in-plane lattice parameter is only $3.7787$ ${\AA}$\cite{lattice_FeSe}. Furthermore, the bands of Li(OH)FeSe disperse weakly along the $k_z$-direction and there is  the band inversion   at the Z point as well. Namely, Li(OH)FeSe is a 3D WTI. Furthermore, the $E_{g+}$ band has angular momentum $\pm\frac{3}{2}$  and  the $A_{2u}$ band has angular momentum $\pm\frac{1}{2}$ so that the WTI phase of Li(OH)FeSe is of the case described by Eq.\ref{H0_WTI2}. As a result, Li(OH)FeSe can only have a narrow nodal superconducting vortex line phase.  If there is a nematic order\cite{nematic1,nematic2,nematic3,nematic4},  as shown in Fig.\ref{iron_phase}(e), the nodal phase can be broken into a full-gap one and vortex end MZMs can emerge on the (001) surface.

Besides the above WTI phase, Ref.\cite{iron_Hao} points out that a 2D "WTI" state, which is attributed to a double-band inversion at the M point, may exist in single-layer FeSe. For the large lattice parameter in the $c$-direction, the band structures of LiOHFeSe are similar to that of the single-layer FeSe. Therefore, we also consider the 2D "WTI" phase here. Based on the symmetry analysis, the bands near the Fermi level at the M point in iron-based SCs are described by the following Hamiltonian in the one-Fe unit cell\cite{iron_kp}
\begin{eqnarray}\label{H0_iron}
H_0^{iron}&=&\left(\begin{array}{cc}
          H_X                 &     H_{inter}     \\
          H_{inter}^\dagger   &     H_Y           \\
          \end{array}
          \right),
\end{eqnarray}
in which $H_X$ and $H_Y$ describe the bands at the X and Y point (defined in the one-Fe unit cell) respectively, and $H_{inter}$ is the inter-pocket hybridization term. The 2D "WTI" phase in Ref.\cite{iron_Hao} corresponds to such a condition that, $H_{inter}$ vanishes and both $H_X$ and $H_Y$ describes 2D TIs, which is similar to the system in Eq.\ref{H0_WTI_2D}. However, $H_X$ and $H_Y$ are related with each other by the $C_{4z}$ rotational symmetry in iron-based SCs. That is to say, $H_X$ and $H_Y$ must have the same critical chemical potential, namely $\mu^c_1=\mu^c_2$, resulting in that the 2D "WTI" phase itself can not lead to topologically nontrivial vortex line states. However, it has been intensively studied that  the $C_{4z}$ rotational symmetry are intended to be broken, namely,   there  can be a nematic order\cite{nematic1,nematic2,nematic3,nematic4} $\Delta_{ne}$ in doped LiOHFeSe, (Li$_{1-x}$Fe$_{x}$)OHFeSe. In this case, the critical chemical potentials for $H_X$ and $H_Y$  are different and  satisfy $|\mu^c_1-\mu^c_2|\propto\Delta_{ne}$. When the nematic order is larger than the inter-pocket hybridization, (Li$_{1-x}$Fe$_{x}$)OHFeSe must have a topologically nontrivial vortex line phase for certain range of doping level. Furthermore, the above topological phase must be a full-gap one, since the system only has $C_{2z}$ rotational symmetry, which means that there can be vortex end MZMs in (Li$_{1-x}$Fe$_{x}$)OHFeSe, as shown in Fig.\ref{iron_phase}(e).

We can extend the above study to  other iron-based SCs.  Recent studies have shown that there is a TDS phase in LiFeAs\cite{iron_TDS}, BaFe$_2$As$_2$\cite{iron_TDS} and FeSe$_{0.5}$Te$_{0.5}$\cite{FeTeSe_3D} with heavy electron doping. In LiFe$_{0.97}$Co$_{0.09}$As, the Dirac point about 20 meV above the Fermi level has been observed  experimentally\cite{iron_TDS}.  The two Dirac points on the $\Gamma$-Z line in LiFeAs are contributed by the band cross between the $d_{xz}/d_{yz}$ and $p_z$ orbitals, whose $z$-component of the angular momentum are $\pm\frac{3}{2}$ and $\pm\frac{1}{2}$ respectively. Consequently, its superconducting vortex line must be nodal if the chemical potential is near the Dirac points. Besides the TDS phase, LiFeAs also has a STI phase near the Fermi level\cite{iron_TDS}. Correspondingly, it must has a topologically nontrivial full-gap vortex line phase when it is in the STI phase. Based on the above results,  the topological phase diagram for superconducting LiFeAs plus a quantum vortex line should be described by  Fig.\ref{iron_phase}(d). Since FeSe$_{0.5}$Te$_{0.5}$ has similar normal state topological properties with LiFeAs\cite{iron_TDS,FeTeSe_3D}, its superconducting vortex line has a similar phase diagram.

For the doped LiOHFeSe,  (Li$_{1-x}$Fe$_{x}$)OHFeSe, besides our above discussion, we also find that the interlayer coupling for the $A_{2u}$ band increases as $x$ increases. Therefore, in principle, there is a critical value, $x=x_c$, at which a topological phase transition from 3DWTI to 3DTDS  can take place.

\section {\bf Discussion and conclusion}

In summary, we study the topological VPTs in iron-based SCs. Besides the previously known phases, we show that there are topoloigcal vortex phases associated with the existence of 3DWTI.  The phases can be classified into two classes: if the band inversions occur on a $C_n$ ($n>2$) rotational axis, and the angular momentums of the orbitals contributing to the band inversions are not equivalent, the superconducting WTI plus a quantum vortex line has a robust 1D nodal superconducting phase with bulk MZMs; otherwise, the superconducting vortex line is full-gap and there can be vortex end MZMs for certain range of doping level. In both cases, the range of the topologically nontrivial phase is proportional to the band dispersion along the weak coupling direction.  All these phases can be realized in different families of iron-based SCs by carreir doping and modifying interlayer couplings.

It should be emphasized that, the nodal superconducting vortex line state can only be distinguished from the full-gap one at rather low temperature, since the Caroli-de Gennes-Matricon excitations in a vortex line generally has a tiny gap proportional to $\Delta^2/E_f$\cite{Caroli}, where $\Delta$ is the superconducting order parameter and $E_f$ is the Fermi energy. For instance, LiFeAs has a $T_c$ about 18 K. If we ignore its topological nontrivial property in normal state, this leads to a full-gap superconducting vortex line with a gap about 0.1 meV when $E_f=20$ meV. Thus, the temperature to detect the nodal superconducting state must be less than 1 K.

From our study, it is clear that the VPTs are sensitive to the chemical potential in superconducting 3D STI and is even more sensitive in a superconducting 3DWTI. This sensitivity suggests that the disorder can have a strong effect on the VPT. Furthermore, as iron-based SCs are layered structures, the interlayer coupling is  weak in general. Considering that the bands in iron-based SCs are known to be renormalized by electron-electron correlations strongly, we can argue that even FeSe/Te can be close to a 3DWTI.  This may explain why MZMs only appear in some vortices\cite{FeSe_feng,TSC_iron2}.


\section {\bf acknowledgement}

S. S. Qin thanks N. N. Hao for helpful discussions. This work is supported by the Ministry of Science and Technology of China 973 program(Grant No. 2014CB921203, No. 2015CB921300, No. 2017YFA0303100), National Science Foundation of China (Grant No. NSFC-11334012, No. NSFC-11674278), and the Strategic Priority Research Program of CAS (Grant No. XDB07000000).



\end{document}